\documentclass[graybox]{svmult}

\usepackage{mathptmx}       
\usepackage{helvet}         
\usepackage{courier}        
\usepackage{type1cm}        
\usepackage{makeidx}         
\usepackage{graphicx}        
\usepackage{multicol}        
\usepackage[bottom]{footmisc}

\usepackage{natbib}
\usepackage{amssymb}
\usepackage{amsmath}

\makeindex             

\newcommand{\msigma}{$M_{\rm BH}-\sigma_{\star}$}
\newcommand{\rl}{$R_{\rm BLR}-L_{\rm AGN}$}
\newcommand\ion[2]{#1$\;${\small\uppercase\expandafter{\romannumeral #2}}\relax}
\newcommand\degr{\mbox{$^\circ$}}%
\newcommand\aj{AJ}%
\newcommand\araa{ARA\&A}%
\newcommand\apj{ApJ}%
\newcommand\apjl{ApJ}%
\newcommand\apjs{ApJS}%
\newcommand\mnras{MNRAS}%
\newcommand\pasp{PASP}%
\newcommand\pasj{PASJ}%
\newcommand\ssr{SSR}
\newcommand\aap{A\&A}

\begin{document}

\title*{AGN Reverberation Mapping}
\author{Misty C.\ Bentz}
\institute{Misty C.\ Bentz \at 
Georgia State University, Department of Physics and Astronomy, 25 Park Place Suite 605, Atlanta, GA 30303, 
\email{bentz@astro.gsu.edu}}

%
%
\maketitle

\abstract*{}

\abstract{Reverberation mapping is now a well-established technique
  for investigating spatially-unresolved structures in the nuclei of
  distant galaxies with actively-accreting supermassive black holes.
  Structural parameters for the broad emission-line region, with
  angular sizes of microarcseconds, can be constrained through the
  substitution of time resolution for spatial resolution.  Many
  reverberation experiments over the last 30 years have led to a
  practical understanding of the requirements necessary for a
  successful program.  With reverberation measurements now in hand for
  60 active galaxies, and more on the horizon, we are able to directly
  constrain black hole masses, derive scaling relationships that allow
  large numbers of black hole mass estimates throughout the observable
  Universe, and begin investigating the detailed geometry and
  kinematics of the broad line region.  Reverberation mapping is
  therefore one of the few techniques available that will allow a
  deeper understanding of the physical mechanisms involved in AGN
  feeding and feedback at very small scales, as well as constraints on
  the growth and evolution of black holes across cosmic time.  In this
  contribution, I will briefly review the background, implementation,
  and major results derived from this high angular resolution
  technique.  }

\section{Introduction and Motivation}
\label{sec:1}

In the 25 years that the {\it Hubble Space Telescope} has been in low
Earth orbit, there have been leaps and bounds in our understanding of
many astrophysical phenomena, not the least of which are supermassive
black holes.  Dedicated ground-based programs coupled with the
exquisite spatial resolution afforded by {\it HST} have led to the
now-common understanding that massive galaxies host supermassive black
holes in their cores (see the review by \citealt{ferrarese05}).  In
general, studies have shown that more massive galaxies host more
massive black holes, which is interpreted as a symbiosis between
galaxies and black holes in which they grow together and regulate each
other's growth.  Recent studies have begun to find several galaxies
that do not appear to follow this simple scaling relationship,
however, and our picture of galaxy and black hole co-evolution
throughout the Universe's history is becoming more complicated (see
the review by \citealt{kormendy13}).

The active galactic nucleus, or AGN, phase is now understood to be a
sporadic event in the life of a typical supermassive black hole,
thought to be triggered by a merger or secular process in the host
galaxy (see the review by \citealt{heckman14} and references
therein). During this phase, the black hole is accreting at a
relatively high rate, and the accretion process is releasing large
amounts of energy across the electromagnetic spectrum.  The lifetime
of a typical AGN event for a typical black hole is small compared to
the age of the Universe, and is generally thought to be on the order
of $\sim 10^8$\,years (e.g., \citealt{kelly10}).  Cosmic downsizing is
observed in AGNs as it is in galaxies: the bright quasars we see in
the early Universe are associated with massive black holes and large
accretion rates, while today's active black holes tend to be fewer in
number, less massive, and have smaller accretion rates (e.g.,
\citealt{ueda03,ueda14,shankar09,kelly10,kalfountzou14}).

Unfortunately, the rarity of bright AGNs compared to the plethora of
galaxies hosting quiescent black holes today leads to the situation in
which we find ourselves, where AGNs are generally distant and
difficult to study even with the pristine spatial resolution afforded
by {\it HST}.  The techniques that have been developed to constrain
the masses of inactive black holes in nearby galaxies rely on
spatially resolving the innermost parsecs of a galactic nucleus, so
they are limited to distances \mbox{$\lesssim 100$\,Mpc} and therefore
not applicable to most AGNs.  And yet AGNs act as beacons that shine
across the entire observable Universe, tempting us to try to
understand black hole and galaxy growth and evolution out to $z=7.1$
\citep{momjian14} and perhaps beyond.

Luckily, AGNs are not only bright, they are also highly variable.  We
can, in effect, substitute time resolution for spatial resolution in a
technique known as reverberation mapping
(\citealt{blandford82,peterson93}) to probe microarcsecond scales in
the nuclei of even the most distant active galaxies.

\section{Reverberation Mapping Primer}
\label{sec:2}

\begin{figure}[t]
\sidecaption[t]
\includegraphics[scale=0.5]{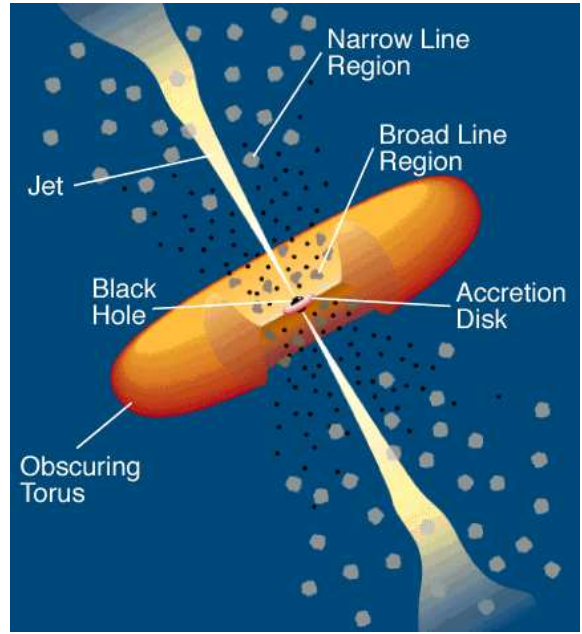}
\caption{Cartoon diagram of the typically-assumed structure of an
  AGN. Depending on the orientation at which this structure is viewed,
  different spectral signatures will be seen.  In particular, the BLR
  is only visible to observers with a relatively face-on view.  If
  viewed from the side, the torus blocks the BLR and only narrow
  emission lines will be seen in the AGN spectrum. From
  \citet{urry95}.}
\label{fig:uni}       
\end{figure}

Many independent studies have led to the general picture of AGN
structure that we understand today (e.g.,
\citealt{antonucci93,urry95,peterson97,netzer15} and references
therein), as represented by the cartoon diagram in
Figure\,\ref{fig:uni}.  In the center is the supermassive black hole,
with a mass in the range of $10^6-10^{10}$\,M$_{\odot}$, and its
associated accretion disk.  The jet (if the AGN has one) is
perpendicular to the accretion disk and highly collimated.  On larger
scales ($\sim 0.01$\,pc for typical Seyferts, approximately the extent
of the inner Oort Cloud in our own Solar System) lies a region of
photoionized gas that radiates line emission. The location of this
gas deep within the potential well of the black hole results in
line-of-sight gas velocities that are quite large, causing the
emission lines to appear Doppler broadened in the AGN spectrum by a
few 1000\,km\,s$^{-1}$.  We imaginatively call this region of gas the
broad line region (BLR).  The outer edge of the BLR is most likely set
by the dust sublimation radius (e.g.,
\citealt{netzer93,nenkova08,goad12}), as the inclusion of dust in the
BLR gas will extinguish the line emission. The dusty gas that exists
outside this radius is generally referred to as the ``torus'',
although the exact geometry of the region is not known.  The dust
torus causes the AGN system to have different spectral signatures when
viewed at different orientations --- a system that is close to face-on
will have broad emission lines in its spectrum, while a system that is
viewed edge-on will have the dust torus blocking the observer's line
of sight to the BLR, so no broad emission lines will be seen.  On even
larger scales ($\sim$tens of pc for typical Seyferts), additional gas
that is photoionized by the AGN system also exists, but the location
of this gas on galactic scales results in line-of-sight velocities
that are on the order of a few 100\,km\,s$^{-1}$.  We see the
signature of this gas as narrow emission lines in the AGN spectrum,
hence the name attributed to this region of gas is the narrow line
region (NLR).

\begin{figure}[t]
\center
\includegraphics[scale=0.5]{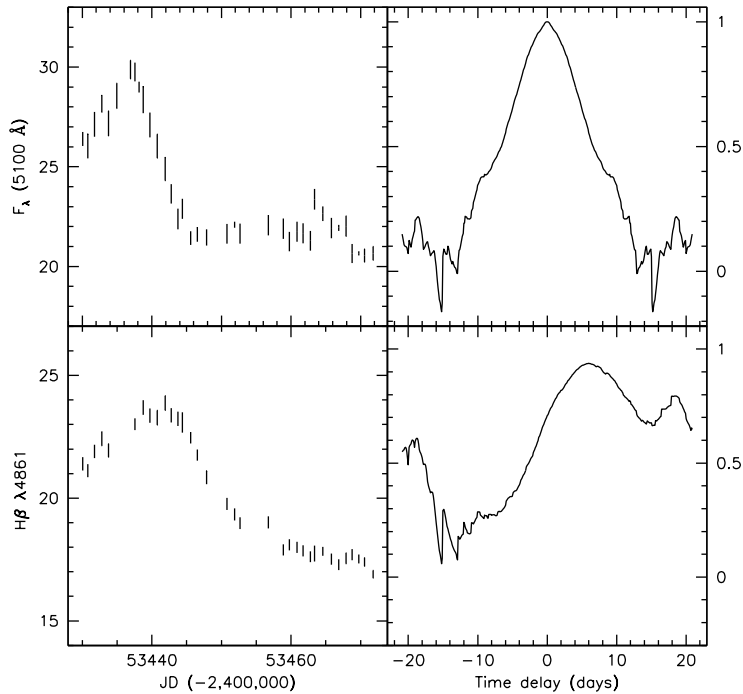}
\caption{Continuum and H$\beta$ light curves for the Seyfert galaxy
  NGC\,4151 ({\it left}), autocorrelation function of the continuum
  light curve ({\it top right}), and cross-correlation of H$\beta$
  relative to the continuum ({\it bottom right}).  The emission-line
  light curve is visibly delayed from that of the continuum and
  smoothed in time, both signatures that are obvious to the eye in the
  light curves and the cross-correlation function, and indicative of
  the extended nature of the H$\beta$-emitting BLR gas. From
  \citet{bentz06}.}
\label{fig:lc}       
\end{figure}

Reverberation, or ``echo'', mapping measures the light travel time
between different regions in an AGN system.  The continuum emission is
expected to arise from the accretion disk, and in the ultraviolet,
optical, and near-IR it is observed to vary on timescales of hours to
days.  The source of this variability is not yet understood (one
possible explanation is magnetic recombination in the accretion disk,
e.g., \citealt{kawaguchi00}), but whatever the cause, the variations
that are observed in the continuum emission are seen echoed at a later
time in the fluxes of the broad emission lines (see
Figure\,\ref{fig:lc} for example light curves for the nearby Seyfert
galaxy NGC\,4151).  The time delay between the variations in the
continuum and the echo of those variations in an emission line is
simply the average light travel time from the accretion disk to the
BLR.  The accretion disk is generally assumed to be very compact, and
so the time delay can be interpreted as the average radius of the BLR
in the AGN system.  By definition, reverberation mapping requires a
line of sight that permits the observer to view the broad emission
lines in the AGN spectrum, so it is only applicable to AGNs with a
relatively face-on orientation.

Within the BLR, different emission lines are observed to respond to
continuum variations with different time delays, such that species
with higher ionization potentials, like \ion{C}{4} $\lambda 1549$,
respond with a shorter time delay than those with lower ionization
potentials, like H$\beta$ (e.g.,
\citealt{peterson00,kollatschny01,bentz10b}).  This behavior points to
ionization stratification within the BLR -- more highly ionized line
emission is radiated from a smaller radius within the BLR, while
emission from more neutral gas occurs at larger radii, further from
the central ionizing source.  Photoionization modeling constraints
agree with this interpretation: the photoionized gas in the BLR
preferentially emits line emission wherever the temperature and
density are most favorable for a specific atomic transition (e.g.,
\citealt{baldwin95,korista97,korista04}).

With reverberation mapping, the radius we measure is the
responsivity-weighted average radius, which may not exactly coincide
with the emissivity-weighted average radius of the gas for that
particular line emission.  But the general behavior that we observe is
the same: in a single AGN, if the central luminosity increases, then
the time delay we measure for any specific emission line becomes
larger (e.g., \citealt{peterson02,bentz07,kilerci15}).  It is
important to keep in mind that we are not measuring a geometrical
radius, such as an inner or outer boundary of the BLR, with
emission-line reverberation mapping.

Even for a relatively nearby AGN, the size of the region probed by
reverberation mapping is quite compact and not resolvable with current
imaging technology or that which is likely to be developed in the
foreseeable future.  For a typical Seyfert galaxy at a distance of
40\,Mpc, the time delay expected for the H$\beta$ $\lambda 4861$
emission line is $\sim 10$\,light-days, which projects to an angular
radius of $\sim 50$\,microarcsec.

In practice, reverberation mapping relies on dense spectrophotometric
monitoring of an AGN system over an extended period of time.  The
basic requirements for a successful monitoring program are the following:
\begin{enumerate}
\item{total campaign length at least three times the longest expected
  time delay to maximize the probability that varibility of a large
  enough amplitude will occur during the campaign \citep{horne04};}
\item{sampling cadence that is sufficiently dense to resolve the
  variability and expected time delays (e.g., Nyquist sampled in
  time);}
\item{exposure times that yield a signal-to-noise $> 50$ in the
  continuum and substantially higher in the emission lines, where the
  amplitude of variability is generally only a few percent;}
\item{spectral resolution that is sufficiently high to distinguish
  broad emission lines from each other and from overlapping or nearby
  narrow emission lines (generally $R$ of a few thousand);}
\item{flux calibration that is good to $2$\% or better from
  observation to observation throughout the campaign
  \citep{peterson04};}
\item{strong nerves and a healthy dose of good luck.}
\end{enumerate} 

Meeting all of these constraints with a ground-based telescope is
observationally quite challenging.  The weather especially can cause
an otherwise well-planned reverberation campaign to not live up to its
potential or to fail outright.  Added to this uncertainty is the fact
that AGN variations are stochastic and not guaranteed to occur during
the course of a monitoring campaign for any particular AGN of interest
(cf.\ the case of Mrk\,290 which was monitored in 2007 and showed
strong variations as reported by \citealt{denney10}, but according to
\citealt{bentz09} showed little to no variation when it was monitored
again in 2008).  The typical warning for the stock market also applies
here: past performance is no guarantee of future behavior.

To date, successful reverberation campaigns have been carried out for
$\sim 60$ different AGNs \citep{bentz15}.  Constraints on
readily-available resources have generally limited the size of the
telescope used in a reverberation campaign to $1.0-4.0$-m class
telescopes.  Coupled with the need for high signal-to-noise ratio
spectra in each visit, most of these 60 AGNs are apparently bright and
reside within the nearby Universe ($z<0.1$).  Reverberation mapping is
not fundamentally limited to nearby objects, but the high luminosities
necessary for high signal-to-noise spectra of $z=2-3$ quasars directly
translates to a long emission-line time delay, which is further
stretched through cosmological dilation (cf.\ \citealt{kaspi07} and
their monitoring campaign length of 10 years).  Ongoing efforts to
multiplex reverberation mapping with the Sloan Digital Sky Survey
multi-object spectrograph may soon increase the sample size by a
substantial fraction and push the median of the sample to somewhat
larger redshifts \citep{shen15}, but large statistical samples of
reverberation results at multiple redshifts spanning the course of
cosmic history are unlikely for the forseeable future.  Luckily, the
results we have in hand for the current reverberation sample are able
to provide us with a foothold for investigating cosmological black
hole growth.

\section{Reverberation Mapping Products}
\label{sec:3}

Substantial progress over the last 10 years especially has led to
several valuable and widely used reverberation-mapping products,
including direct black hole mass measurements, black hole scaling
relationships that can be used to quickly estimate large numbers of
black hole masses, and detailed information on the geometry and
kinematics of the BLR gas.  As we describe in the following sections,
the first two items are fairly well-developed at this time, while we
are just now beginning to truly exploit the third.

\subsection{Black Hole Masses}
\label{sec:3.1}

In order to directly constrain the mass of a black hole, a luminous
tracer (usually gas or stars) must be used to probe the local
gravitational potential.  In the case of reverberation mapping, the
photoionized BLR gas is deep within the potential well of the black
hole where its motion is expected to be dominated by gravity in the
absence of strong radiation pressure.  While the effect of radiation
pressure is still debatable for lines such as \ion{C}{4}, these
conditions are most likely to be satisfied for H$\beta$ (e.g.,
\citealt{marconi08,netzer09,netzer10}) which is expected to arise from
gas that is well-shielded from the central ionizing source.  The
Doppler-broadened width of the H$\beta$ emission line is therefore a
measure of the line-of-sight velocity of the gas within the BLR.  And,
as described in Section\,\ref{sec:2}, the time delay in the H$\beta$
emission line is a measure of the radius of the BLR for that same
line-emitting gas.  Coupling these two measurements together through
the virial theorem allows a direct constraint on the mass of the
central black hole, modulo a scaling factor to account for the
detailed geometry (including inclination) and kinematics (whether
rotation, infall, or outflow) of the gas.

Given the orientation-dependent picture of AGNs described above, and
our inability to directly resolve the BLRs of even nearby AGNs with
current technology, it is necessary to constrain the black hole mass
scaling factor through some means.  This is typically accomplished in
an indirect way by comparing the relationship between black hole mass
and host-galaxy bulge stellar velocity dispersion, the
\msigma\ relationship, for nearby quiescent galaxies with stellar- and
gas-dynamical modeling-based black hole masses to the relationship for
AGNs with as-yet-unscaled reverberation masses, i.e., virial products.
The average multiplicative factor that must be applied to the AGN
virial products to bring the AGN relationship into agreement with that
of the quescent galaxies is found to be $\sim 4-5$ (most recently $4.3
\pm 1.1$, \citealt{grier13}, when all AGNs are treated equally and
local high-luminosity AGNs are included to extend the range of the
\msigma\ relationship).

Certainly, there are many assumptions involved in determining this
average scaling factor, and it is important to keep in mind when
applying the scaling factor that it is a population average and
therefore likely to be uncertain by a factor of $2-3$ for any
individual AGN. Nonetheless, several lines of independent evidence
point to this value of the mass scaling factor being in the right
ballpark.

The first is to simply assume that the main contribution to the scale
factor is the inclination of the system, or the $\sin i$ term in the
gas velocity.  A scale factor of $4.3$ would imply a typical
broad-lined AGN inclination of $\sim29$\degr, which is well in line
with expectations based on our current understanding of AGN structure.
Furthermore, this average inclination agrees well with the geometric
inclinations derived for the extended and resolved narrow-line region
structure of several nearby AGNs \citep{fischer13}.

\begin{figure}[t]
\sidecaption[t]
\includegraphics[scale=0.35]{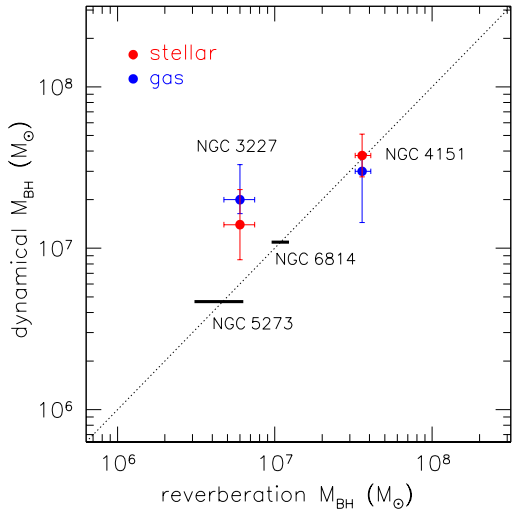}
\caption{Comparison of black hole masses derived from reverberation
  mapping with an average mass scale factor applied, and from stellar
  dynamical modeling and gas dynamical modeling which do not rely on a
  scale factor.  Stellar dynamical modeling is in progress for
  NGC\,6814 and NGC\,5273, but the expected location based on the
  reverberation mass for each is denoted in the figure.}
\label{fig:masscomp}       
\end{figure}

A more rigorous test is to compare the black hole masses derived from
reverberation mapping with those derived from stellar- or
gas-dynamical modeling in the same objects.  As previously discussed,
most AGNs are too far away for dynamical modeling techniques to be
applied, but a few very nearby AGNs can be examined in this way.
Direct comparisons have been carried out for two AGNs to date ---
NGC\,3227 \citep{davies06,denney10} and NGC\,4151 \citep{onken14} (see
Figure\,\ref{fig:masscomp}).  So far, the resultant black hole masses
agree remarkably well between such disparate measurement techniques,
each derived from independent observations and each with their own
independent set of assumptions and biases.  While the agreement is
reassuring for these two objects, a sample of two is hardly
definitive. Two more AGNs with reverberation masses --- NGC\,6814
\citep{bentz09} and NGC\,5273 \citep{bentz14} --- are in various
stages of the dynamical modeling process, and a handful of other AGNs
are being targeted for reverberation-mapping with the hope of
dynamical modeling to follow.  

One current complication to this test, however, is the effect of bars
on the central stellar dynamics, and therefore the derived black hole
masses (e.g., \citealt{brown13,hartmann14}).  Even for the relatively
face-on galaxy NGC\,4151, \citet{onken14} found a significant bias was
induced in the best-fit black hole mass from the weak galaxy scale
bar.  NGC\,5273 will be an especially interesting case for testing
black hole masses from reverberation mapping versus stellar dynamical
modeling given its unbarred S0 morphological type.  Furthermore, in
the next several years, it is likely that JWST will allow some
advancement in the numbers of AGNs that can provide direct mass
comparisons across techniques.  \citet{gultekin09} argue that
dynamical modeling can still place strong constraints on black hole
mass even if the radius of influence of the black hole is not strictly
resolved in the observations.  JWST will provide a comparable spatial
resolution for studies of host-galaxy stellar absorption features to
that which is currently achieved with ground-based observatories and
adaptive optics, but the advantages of JWST include a stable PSF, a
significantly higher Strehl ratio, and very low backgrounds, all of
which are important for deriving tight dynamical constraints on the
black hole mass.

Finally, as we discuss in Section\,\ref{sec:3.3}, it is possible to
directly constrain the black hole mass, without needing to resort to
the use of a scaling factor, from the reverberation-mapping data
itself.  Data quality concerns have generally not allowed this goal to
be met in the past, but recent progress is encouraging, and the
resultant mass constraints generally agree with our expectations based
on the arguments above.

\subsection{Black Hole Scaling Relationships}
\label{sec:3.2}

\begin{figure}[t]
\center
\includegraphics[scale=0.45]{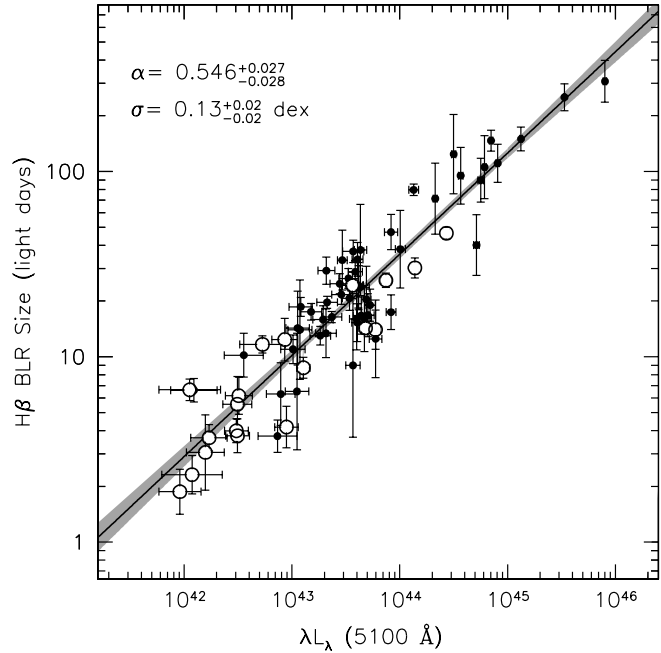}
\caption{The relationship between the H$\beta$ time delay and the
  specific luminosity of the AGN at 5100\,\AA, the \rl\ relationship.
  From \citet{bentz13}.}
\label{fig:rl}       
\end{figure}

One of the most useful scaling relationships to arise from the
compendium of reverberation-mapping measurements is the
\rl\ relationship --- the relationship between the time delay, or
average radius of emission, for a specific emission line and the
luminosity of the central AGN at some particular wavelength.  This
particular scaling relationship was expected from simple
photoionization arguments and looked for in the early days of
reverberation mapping when the number of successful monitoring
campaigns and reverberation measurements was still very small
\citep{koratkar91}.  The addition of several high-luminosity local PG
quasars to the reverberation sample led to the first well-defined
functional form of the \rl\ relationship \citep{kaspi00}.
Figure\,\ref{fig:rl} shows the most current calibration of the
\rl\ relationship between H$\beta$ and the AGN continuum luminosity at
rest-frame 5100\,\AA, where all luminosities have been carefully
corrected for the contamination of host-galaxy starlight using
high-resolution {\it HST} images and two-dimensional decompositions of
the images to separate the galaxy starlight from the AGN emission
\citep{bentz13}.

Rather than carrying out a long-term monitoring campaign for any AGN
of interest, the \rl\ relationship allows a single spectrum to provide
an estimate of the black hole mass through two simple measurements:
the width of the broad emission line as a proxy for the gas velocity,
and the continuum luminosity of the AGN as a proxy for the time delay
expected in the emission line.  This handy shortcut provides a means
for taking large spectroscopic surveys and producing catalogs full of
black hole mass estimates (e.g., \citealt{shen11}).  Of course, as one
might expect, the devil is in the details.

Currently, H$\beta$ is the only emission line for which a
well-calibrated \rl\ relationship exists.  Unfortunately, H$\beta$
shifts out of the observed-frame optical bandpass at redshifts of only
$z=0.7-0.8$.  The \ion{Mg}{2} and \ion{C}{4} emission lines in the
rest-frame ultraviolet are therefore more accessible for most quasar
surveys conducted from the ground.  But at this time, only a handful
of measurements of \ion{C}{4} reverberation time delays exist (see
\citealt{kaspi07} for a first attempt to constrain a \ion{C}{4}
\rl\ relationship), and even fewer reverberation measurements exist
for \ion{Mg}{2}.  Quasar black hole masses therefore require
bootstrapping the estimates into the UV using the H$\beta$
\rl\ relationship as the cornerstone (e.g., \citealt{vestergaard06}).
Furthermore, the exact prescription for turning two simple spectral
measurements into an unbiased mass estimate is still highly debatable
(see, e.g., \citealt{denney09,denney13} for discussion of several of
the specific details that can cause biases).

Interestingly, the small scatter in the \rl\ relationship has led to
the proposal that it may be used to turn AGNs into standardizable
candles for investigating cosmological expansion \citep{watson11,king14}.
Quasars are easily observed well beyond $z \approx 1$, where Type Ia
supernovae become rare and difficult to find and where the differences
in cosmological models are more apparent.  One of the largest
practical difficulties in turning this idea into reality, however, is
again the long time delays involved in monitoring high redshift
quasars and the necessity of high signal-to-noise spectroscopy over
the course of such a monitoring campaign.

Other black hole scaling relationships include the aforementioned AGN
\msigma\ relationship \citep{onken04,graham11,park12,grier13} and the
relationship between the AGN black hole mass and the host-galaxy bulge
luminosity, the $M_{\rm BH} - L_{\rm bulge}$ relationship
\citep{wandel02,bentz09b,bentz11}.  While the more commonly-used forms
of these relationships tend to be those derived for black holes with
dynamically-modeled black hole masses (e.g.,
\citealt{magorrian98,ferrarese00,gebhardt00,gultekin09,mcconnell13,kormendy13}),
the AGN relationships provide a useful counterpoint given the
differences between the two samples.  In particular, the AGN
reverberation sample has a large percentage of late-type,
disk-dominated galaxies, whereas the quiescent galaxy sample with
dynamical black hole masses is comprised mainly of early-type
galaxies.  It is not at all clear that galaxies of different
morphological types should all follow the same scaling relationships
(cf.\ the recent review by \citealt{kormendy13}).  Furthermore, active
galaxies may not follow the same scaling relationships as quiescent
galaxies (e.g., \citealt{wandel99}), but most studies of galaxy and
black hole co-evolution at cosmological distances are necessarily
limited to active galaxies.

Even despite these differences between the samples, it appears that
our current constraints on the general forms of the $M_{\rm BH} -
L_{\rm bulge}$ and \msigma\ relationships are consistent for AGNs with
reverberation masses and for quiescent galaxies with dynamical black
hole masses.  Several ongoing studies aim to refine and more
accurately constrain these relationships, and part of this effort is
devoted to tackling the key observational uncertainties that remain
--- such as determining accurate distances to the AGN host galaxies,
and replacing stellar velocity dispersion measurements from long-slit
spectra with those obtained from integral field spectroscopy of the
host galaxies.  Here again, barred galaxies (and galaxies with
``disky'' bulges) are a source of confusion.  While they are seen to
be outliers in the quiescent \msigma\ relationship
\citep{hu08,graham11,kormendy13}, there is no such offset seen in the
AGN \msigma\ relationship \citep{grier13} unless it is artificially
inserted by scaling the black hole masses in those galaxies by a
different value \citep{ho14}.  Comparisons between the different
assumptions and biases in the AGN \msigma\ relationship versus the
quiescent galaxy relationship will therefore help to shed light on the
underlying causes for such puzzles. And while it has been the focus of
less intense study in the last decade or so, an accurately calibrated
$M_{\rm BH} - L_{\rm bulge}$ relationship will be especially necessary
for constraining galaxy evolution through upcoming deep all-sky
photometric surveys with no spectroscopic component, like LSST.

\subsection{BLR Geometry and Kinematics}
\label{sec:3.3}

\begin{figure}[b]
\includegraphics[scale=0.4]{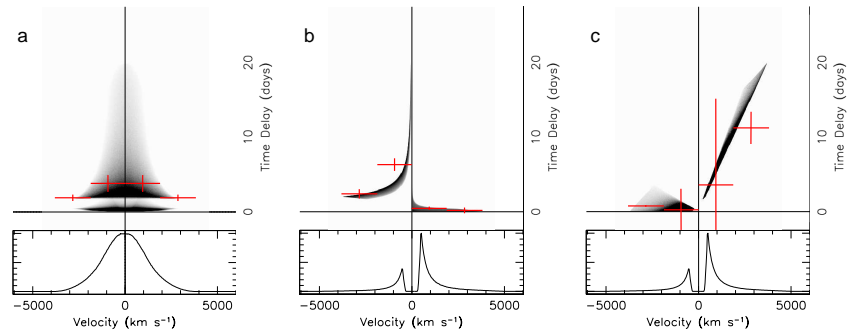}
\caption{The expected emission-line response for a toy model BLR with
  three different possible kinematics: (a) rotation, (b) infall, and
  (c) outflow.  For simplicity, the geometry is kept the same for all
  three cases --- the line emission is restricted to a bicone with a
  semi-opening angle of 30\degr\ and the model is inclined at 20\degr\
  so that the observer is inside the beam. The radiation structure
  within the BLR clouds is set so that the emission is enhanced for
  clouds at smaller radii, and the line emission is partially
  anisotropic, such that the emission is enhanced in the direction of
  the illuminating source.  The gray-scale images show the full
  two-dimensional structure in time lag versus line-of-sight velocity,
  while the vertical red error bars show the weighted mean and
  standard deviation of the time lag within discrete velocity bins
  that are represented by the horizontal red error bars.  The overall
  shape is different for each of the three models: a symmetric
  structure around zero velocity for circular Keplerian orbits, longer
  lags in the blueshifted emission for infall, and longer lags in the
  redshifted emission for outflow.  From \citet{bentz09}.}
\label{fig:simple}       
\end{figure}

Most of the progress in reverberation mapping has focused on the
lowest-order measurement that can be made, namely the average time
delay of an emission line, because this is what is required to make a
dynamical mass measurement.  Historically, it was also the only
measurement that could be recovered from the marginally-sampled light
curves from early reverberation campaigns.  However, there is much
more information encoded in densely-sampled light curves.  In
particular, the emission-line light curve is a convolution of the
continuum variations and the extended response of the BLR gas at
different line-of-sight velocities and light travel times relative to
the observer.  Resolving the time delays as a function of velocity
across an emission-line profile can therefore give constraints on the
detailed geometry and kinematics of the BLR gas.  In
Figure\,\ref{fig:simple}, we show three examples of emission-line
response given one fairly simple model for the BLR and three different
possible kinematics of the gas: rotation, infall, and outflow.  The
differences between the three are apparent in a full deconvolution of
the emission-line response (Figure\,\ref{fig:simple} shaded regions),
or a first-order analysis in which the mean time delays are computed
for velocity bins across the line profile (Figure\,\ref{fig:simple}
error bars).

While the wealth of information that is potentially available from
reverberation-mapping datasets has been understood for quite some
time, the practical difficulties involved in deconvolving a faint
signal from sparsely- and irregularly-sampled noisy data have limited
much progress in this area.  Notable early attempts include
\citet{wanders95}, \citet{done96}, and \citet{ulrich96}, but recovered
maps of velocity-resolved results, reminiscent of those shown in
Figure\,\ref{fig:simple}, were ambiguous at best.  In the last few
years, however, reverberation programs have enjoyed much more success
given the careful experimental setup and the resultant high quality of
the data (\citealt{bentz09,bentz10,denney10,grier12,derosa15}).

With such high quality data now in hand, there are two general methods
for extracting the reverberation signal.  Deconvolution techniques can
be used to produce a model-independent, but potentially
difficult-to-interpret, velocity delay map like the models displayed
in Figure\,\ref{fig:simple}.  Direct modeling of the spectroscopic
data, on the other hand, produces easily-interpreted constraints on
different possible physical models, but is by definition
model-dependent and human imagination-limited.

\begin{figure}[t]
\includegraphics[scale=0.45]{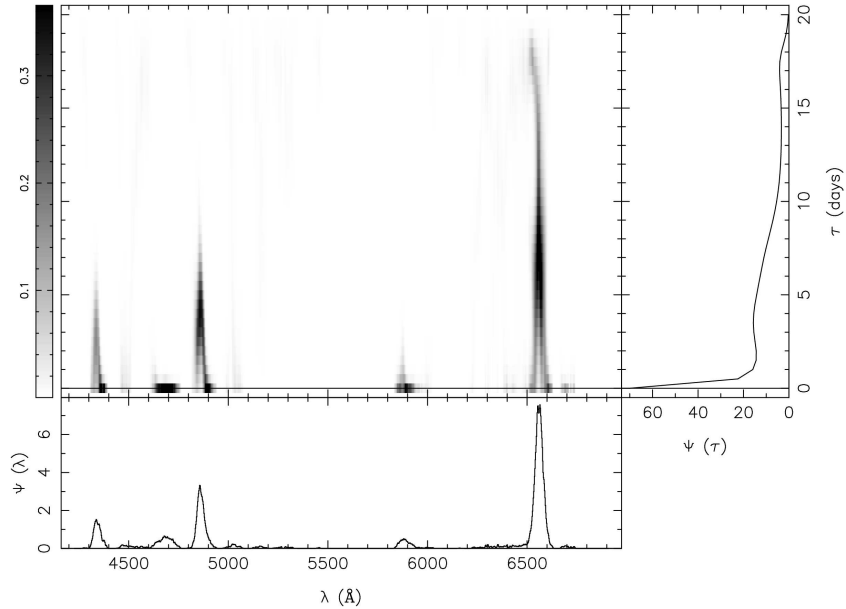}
\caption{The deconvolved emission-line response as a function of
  velocity for the broad optical recombination lines in the spectrum
  of Arp\,151.  From \citet{bentz10}.}
\label{fig:velres}       
\end{figure}

The most widely employed deconvolution algorithm to date is the
MEMECHO code (\citealt{horne91,horne94}), which uses maximum entropy
balanced by $\chi^2$ to find the simplest possible solution that fits
the data.  Each pixel in the AGN spectrum can be treated as a separate
velocity bin for which an entire light curve exists throughout the
monitoring campaign.  The code solves for the time delay response
function in all the individual light curves for each of the pixels in
the spectrum.  From the many reponse functions, a map of time delay as
a function of velocity is reconstructed.

Figure\,\ref{fig:velres} shows an example of the deconvolved response
of the broad optical recombination lines in the spectrum of the
Seyfert galaxy Arp\,151 \citep{bentz10}.  The differences in expected
mean time delays for the lines is immediately apparent, and also
interesting is the strong prompt response in the red wings of the
Balmer lines compared to the lack of prompt response in the blue wings
of those same emission lines.  An asymmetric response such as this
could be produced by either rotating gas with enhanced emission in one
location (such as a hot spot or warp), or by infalling gas, or some
combination of these simplistic models.  Similar asymmetries are also
seen in the deconvolved responses of a handful of additional AGNs
\citep{grier13b}.


Direct modeling, on the other hand, tests the data against specific
geometric and kinematic models to constrain the family of models that
best represent the observations.  For the Arp\,151 dataset above,
direct modeling results prefer a thick disk BLR geometry, inclined at
$\sim25$\degr\ to the observer's line of sight, and the kinematics are
dominated by inflow with some contribution from rotation
\citep{pancoast14}, in general agreement with the interpretation of
the deconvolution results.  \citet{pancoast14} also find similar
results for a few additional AGNs.

\begin{figure}[t]
\sidecaption[t]
\includegraphics[scale=0.48]{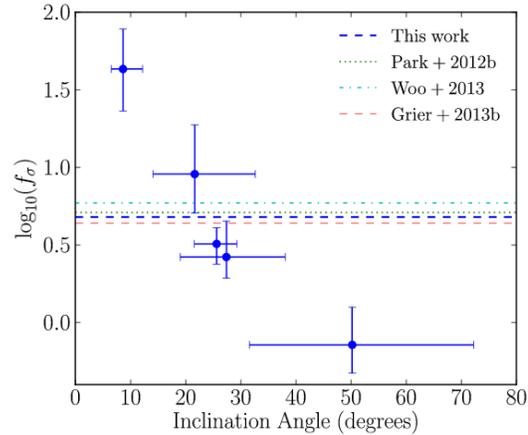}
\caption{ The individual mass scale factor for 5 AGNs with direct
  modeling of their velocity-resolved responses.  The average value
  for this small sample is $f \approx 4.8$ (denoted by the horizontal
  blue dashed line), in good agreement with values determined by
  comparison of the active galaxy and quiescent galaxy
  \msigma\ relationships (denoted by the other horizontal lines).
  From \citet{pancoast14}.}
\label{fig:f_ind}       
\end{figure}

Furthermore, direct modeling is able to determine the {\it individual}
scaling factor that would need to be applied to the
reverberation-based black hole mass as described in
Section\,\ref{sec:3.1} above.  For the handful of AGNs with successful
dynamical models, we can see that the scale factor indeed varies from
object to object, as expected for a population of objects with random
inclinations within some limited range (see Figure\,\ref{fig:f_ind}).
The average of these individual scaling factors also agrees quite well
with the population average derived above through use of the
\msigma\ relationship (see Section\,\ref{sec:3.1}).

\section{Looking Ahead}

The future is looking quite busy for applications of reverberation
mapping.  In addition to the many ongoing areas of study summarized in
the previous pages, several recent or ongoing programs, such as the
multi-object SDSS reverberation program \citep{shen15} and the massive
HST plus ground-based reverberation program for NGC\,5548
\citep{derosa15}, are just starting to report results that are sure to
lead to new insights and new puzzles in AGN physics.  Additionally,
the upcoming OzDES program \citep{king15} will help push to higher
redshifts, providing a more stable anchor for black hole mass
estimates of high-$z$ quasars.

The flurry of recent activity in velocity-resolved reverberation
mapping is unlikely to abate any time soon, and here we may hope to
unlock many of the secrets surrounding AGN feeding and feedback.  New
codes to deal with velocity-resolved reverberation mapping data are
currently being developed (\citealt{skielboe15}, Anderson et al.\, in
prep), and new features are being added to currently-existing codes
(Pancoast, private communication).

With UV astronomy currently dependent on the continuation of {\it
  HST}, it is certainly conceivable to think that the {\it Kronos}
spacecraft \citep{peterson03}, or a similar instrument, may again make
an appearance in proposal form.  It is an exciting time for
supermassive black holes, reverberation mapping, and AGN physics!

\begin{acknowledgement}
I would like to thank the conference organizers for inviting me to
present this review.  This work is supported by NSF CAREER grant
AST-1253702.
\end{acknowledgement}


\end{document}